\documentclass[aps,prl,footinbib,twocolumn]{revtex4-1}
\usepackage{amsmath}
\usepackage{amssymb}
\usepackage{epsfig}
\usepackage{graphics}
\usepackage{bm}
\usepackage{xcolor}

\newcommand{\beq}{\begin{equation}}
\newcommand{\eeq}{\end{equation}}
\newcommand{\bea}{\begin{eqnarray}}
\newcommand{\eea}{\end{eqnarray}}

\def\bp{{\bf p}}
\def\bq{{\bf q}}
\def\br{{\bf r}}
\def\bv{{\bf v}}

\def\calZ{\mathcal{Z}}
\def\calD{\mathcal{D}}
\def\calN{\mathcal{N}}
\def\calA{\mathcal{A}}

\def\calV{\mathcal{V}}
\def\calE{\mathcal{E}}

\def\e{\epsilon}
\def\ve{\varepsilon}

\def\nn{\nonumber}

\begin{document}
\title{Topological superfluid in a Fermi-Bose mixture with a high critical temperature}
\author{Zhigang Wu}
\affiliation{Department of Physics and Astronomy,  Aarhus University, Ny Munkegade, DK-8000 Aarhus C, Denmark}
\author{G.\ M.\ Bruun}
\affiliation{Department of Physics and Astronomy,  Aarhus University, Ny Munkegade, DK-8000 Aarhus C, Denmark}

\date{\today}
\begin{abstract}
We show that a two-dimensional (2D) spin-polarised Fermi gas immersed in a 3D Bose-Einstein condensate (BEC) constitutes a very promising system to realise a $p_x+ip_y$   superfluid. 
The fermions attract each other via an induced interaction mediated by the bosons, and  the resulting pairing is analysed with retardation effects fully taken into account. This is further combined with Berezinskii-Kosterlitz-Thouless (BKT) theory to obtain reliable results for the superfluid critical temperature.
 We show that both the strength and the range of the induced interaction can be tuned experimentally, which can be used to make the critical temperature  approach the  maximum value allowed by  general 
BKT theory. Moreover, this is achieved while keeping the Fermi-Bose interaction weak so that three-body losses are small. Our results show
 that realising a topological superfluid with atomic Fermi-Bose mixtures is  within experimental reach. 
\end{abstract}


\maketitle
The quest for realising topological phases  of matter is presently a very active research topic~\cite{Hasan2010,Qi2011}. Topological superconductors/superfluids are of particular interest, as they exhibit Majorana edge modes with possible applications for quantum computing~\cite{Nayak2008}.
In condensed matter systems, experimental evidence for Majorana modes have  been reported in a number of one-dimensional 
materials~\cite{Mourik2012,Deng2012,Das2012,Rokhinson2012,Finck2013,NadjPerge2014,Albrecht2016}. Furthermore, Sr$_2$RuO$_4$ is believed to realise a 2D topological superconductor~\cite{Ishida1998,Nelson2004,Kidwingira2006,Xia2006}. 
However, it is highly desirable to find other systems which allow for unambiguous realisation of a topological superfluid. 
Cold atomic gases are attractive candidates for this task, as they are devoid of impurities and are highly controllable. 
Fermi gases interacting via a $p$-wave resonance have been suggested to realise topological 
superfluids~\cite{Gurarie2007}, but they are found to have very short lifetimes~\cite{Gunter2005,Gaebler2007,Levinsen2007,JonaLasinio2008}.
Other suggestions using quantum gases include proposals based on optical lattices~\cite{Buhler2014,Massignan2010,Mathey2006,Wu2012}, 
synthetic spin-orbit coupling~\cite{Zhang2008,Sato2009,Jiang2011}, driven dissipation~\cite{Bardyn2012,Diehl2011}, dipolar molecules~\cite{Cooper2009} and 
mixed dimension Fermi-Fermi mixtures~\cite{Nishida2009}. However, none of these proposals have been implemented experimentally so far, partly due to the prohibitively low superfluid critical temperatures in these systems.  

In this paper we show that a 2D-3D Fermi-Bose mixture can realise a $p_x+ip_y$ topological superfluid with a high critical temperature. Spin-polarised 
fermions are confined to a 2D plane and interact via an attractive induced interaction mediated by 
density fluctuations in a weakly interacting 3D BEC.  The superfluid transition of the 2D Fermi gas is investigated by first solving the mean-field gap equation, which has the Eliashberg form due to the frequency dependence of the induced interaction. Here the retardation effects are fully included and are found to be important. The superfluid density is then calculated and the critical temperature of the transition determined using BKT theory. To our knowledge, such a microscopic theory for the pairing of a Fermi gas based on a 
 combination of Eliashberg and BKT theory has not been presented before in the literature. 
We further demonstrate that the strength as well as the range of the induced interaction can be controlled experimentally. This tunability can be utilised to increase the critical temperature of the superfluid transition to the limiting value imposed by BKT theory. Importantly, this is achieved while keeping both the Fermi-Bose and the Bose-Bose interactions weak, which is necessary to minimise three-body losses. Our results suggest a roadmap for realising a long lived topological superfluid in atomic gases with a high critical temperature. 

\emph{Model.--}
We consider a single layer of spin-polarised, non-interacting fermions with mass $m_F$ and areal density $n_F$, immersed in a uniform, weakly interacting 3D Bose gas with particle mass $m_B$ and density $n_B$.
 The interaction between the  fermions and the  bosons is modelled by $g\delta(\br)$. Here the coupling strength is given by $g=2\pi a_{\rm eff}/\sqrt{m_rm_B}$, where $m_r = m_Fm_B/(m_F+m_B)$ is the reduced mass and $a_{\rm eff}$ is the effective 2D-3D scattering length; the latter can be tuned to arbitrary values in atomic gases~\cite{Nishida2008}.  The 2D Fermi gas is assumed to have no effect on the Bose gas, as it is much smaller in size. In the temperature regime relevant to our study, the weakly interacting Bose gas forms a BEC and is well described by the Bogoliubov theory. The partition function of the  2D-3D mixture at temperature $T$ is $(\hbar=k_B=1)$ 
\beq
\calZ = \int  \calD(\bar a, a) \int \calD (b^*,b) e^{-S(\bar a, a;b^*,b)},
\label{pf1}
\eeq
where $(a,\bar a)$ and $(b,b^*)$  are Grassmann and  complex fields
describing the fermions and bosons respectively. The action consists of   
$S(\bar a, a;b^*,b) = S^0_{F} +S_B + S_{\rm int}$.  Here 
$S^0_{F} = \beta\sum_{\bp_\perp,n}  \bar a(p_\perp)\left (-i\omega_n +\xi_{\bp_\perp} \right )a(p_\perp)$ describes free fermions with in-plane momentum 
$\bp_\perp=(p_x,p_y)$, where $\beta=1/T$, $\omega_n =(2n+1)\pi /\beta$ is a Fermi Matsubara frequency and $\xi_{\bp_\perp} = {\bp_\perp^2}/{2m_F} -\mu$ is the fermion spectrum relative to the chemical potential $\mu$ of the Fermi gas.  We have defined  $p_\perp \equiv (\bp_\perp,i\omega_n)$. Within Bogoliubov theory, the action for the 
 BEC is $S_{B} = \beta\sum_{\bp\neq0,\nu}  \gamma^*(p)(-i\omega_\nu+E_\bp)\gamma(p)$. Here $p\equiv (\bp,i\omega_\nu)$ with $\bp=(p_x,p_y,p_z)$, $\omega_\nu =2\nu\pi /\beta $ is a Bose Matsubara frequency, $E_{\bp} =\sqrt{ \e_{\bp} (\e_{\bp} + 2 g_Bn_B) }$ with $\e_{\bp} = \bp^2/2m_B$ is the Bogoliubov spectrum and $\gamma(p) = u_\bp b(p) + v_\bp b^*(-p) $ are the quasi-particle fields. The Bogoliubov amplitudes are $u_{\bp}, v_{\bp} =\sqrt{\frac{1}{2} \left[ \left ({\e_{\bp} + g_Bn_B}\right)/{E_{\bp}} \pm 1 \right]}$, where $g_B = 4\pi a_B/m_B$ with $a_B$ being the boson scattering length. Finally, the Fermi-Bose interaction is given by 
\begin{align}
S_{\rm int}& = g\sqrt{\frac{n_B}{\calV}}\beta\sum_{ \bp\neq 0,\nu}\left [b^*(p) + b(-p) \right ] \rho(p_\perp) \nn \\
&= g\sqrt{\frac{n_B}{\calV}}\beta\sum_{ \bp\neq 0,\nu}\sqrt{\frac{\e_{\bp}}{E_\bp}}\left [\gamma^*(p) + \gamma(-p) \right ] \rho(p_\perp),
\label{Sint1}
\end{align}
 where $\calV$ is the volume of the BEC and  
 $\rho(q_\perp)= \sum_{\bp'_\perp,m} \bar a(\bp'_\perp-\bq_\perp,i\omega_m-i\omega_\nu) a (\bp'_\perp,i\omega_m)$. In Eq.~(\ref{Sint1}) we have
 ignored a term describing scattering between the fermions and the bosons that are not in the condensate, since we focus on the 
 weak Fermi-Bose interaction regime where effects of this term are negligible.   In the case of a strong interaction, however, it 
is crucial to include such a term~\cite{Christensen2015}. We note that the interaction in Eq.~(\ref{Sint1}) does not conserve momentum of two scattering particles along $z$-direction due to confinement of the fermions.

\emph{Induced interaction.--} Performing the integration over the boson fields in Eq.~(\ref{pf1}) we find the following effective action for the fermions
\begin{align}
S_F(\bar a, a) &= S^0_{F}(\bar a, a) +\frac{\beta}{2{\mathcal{A}}}\sum_{q_\perp}V_{\rm ind}(q_\perp)\bar\rho(q_\perp) \rho(q_\perp),
\end{align}
where $\mathcal{A}$ is the area of the  Fermi gas and 
\beq
V_{\rm ind}(\bq_\perp,i\omega_\nu) = \frac{g^2n_B}{m_B}\int_{-\infty}^\infty\frac{dq_z}{2\pi}\frac{q^2}{(i\omega_\nu)^2-E_\bq^2}
\label{Vinddef}
\eeq
is the induced interaction between the fermions mediated by the bosons.  Apart from the additional integration over the $z$-component of the Bose momentum, the formula in Eq.~(\ref{Vinddef}) is similar in form to the induced interaction in a 3D-3D Fermi-Bose mixture~\cite{Heiselberg2000,Bijlsma2000}.
It follows from  
 Eq.~(\ref{Vinddef})  that $V_{\rm ind}(\bq_\perp,i\omega_\nu) $ is manifestly real and negative. Performing the integral in Eq.~(\ref{Vinddef}) we find
\begin{align}
 V_{\rm ind}(\bq_\perp,i\omega_\nu) &= -n_Bm_Bg^2\left [\left(\frac{1}{\kappa_+}+\frac{1}{\kappa_-}\right ) \right. \nn \\ 
 &\left. + \frac{1}{\sqrt{1-(\omega_\nu/g_Bn_B)^2}}\left(\frac{1}{\kappa_+}-\frac{1}{\kappa_-}\right ) \right ], 
 \label{Vindf}
 \end{align}
 where $ \kappa_\pm = \sqrt{2m_Bg_Bn_B\left[1\pm \sqrt{1- (\omega_\nu/g_Bn_B)^2}\right]+\bq_\perp^2}$. Here $\sqrt {z}$ denotes the root of the complex number $z$ with a positive real part. The frequency dependence of the induced interaction reflects the fact that  density oscillations in the BEC have a finite speed.
The most important frequency for pairing is on the order of the Fermi energy $\varepsilon_F=k_F^2/2m_F$, and
  $\varepsilon_F/g_Bn_B \sim (m_F/m_B)(v_F/c_0)^2$, where $v_F=k_F/m_F$ is the Fermi velocity and $c_0=\sqrt{g_Bn_B/m_B}$ is  the speed of sound in the BEC. This suggests that the frequency dependence of the induced interaction can be neglected only if $v_F/c_0\ll 1$, i.e.\ when 
Bogoliubov phonons move at a much greater speed than the fermions. In this case we can set $\omega_\nu=0$ in Eq.\ (\ref{Vindf}), and the induced interaction assumes the simple form $
V_{\rm ind}(\bq_\perp) \simeq -{2n_Bm_Bg^2}/{\sqrt{\bq_\perp^2+2/\xi_B^2}}$, where $\xi_B=1/\sqrt{8\pi n_Ba_B}$ is the BEC coherence length. We point out that the induced interaction in Eq.~(\ref{Vindf})  
is much stronger than that in a 2D-3D Fermi-Fermi mixture with comparable physical parameters~\cite{Nishida2009}, due to the fact that
a BEC is  in general more compressible than a Fermi gas. 

\emph{$p_x+ip_y$ pairing.--}
The attractive induced interaction gives rise to pairing between the  fermions. To describe this, we introduce a pairing field $\Xi(p)$ via the standard 
Hubbard-Stratonovich transformation, which couples  the Grassmann fields with frequency/momenta $p$ and $-p$.
 Since all momenta are now 2D, the $\perp$ sign previously used to distinguish 2D vectors will be dropped henceforth. 
Integrating out the Grassmann fields one finds
\begin{align}
\calZ &= \int \calD(\Xi^*, \Xi) e^{-S_{\rm eff}(\Xi^*,\Xi)}.
\label{Zeff}
\end{align}
Here the effective action is
\begin{equation}
S_{\rm eff}= -\frac{1}{2}{ {\text{Tr}}}\ln G^{-1} -  \frac{\beta\mathcal A }{2}\sum_{p,p'} \Xi^*(p) V_{\rm ind}^{-1}(p-p')\Xi (p')
\label{Seff}
\end{equation}
and $G^{-1}(p)$ is the inverse Green's function
\begin{equation}
G^{-1}(p)= \beta \begin{bmatrix}
    - i\omega_n+\xi_\bp e^{i\omega_n0^+}    &\frac{ \Xi(p)- \Xi(-p)}{2}  \\
     \frac{\Xi^*(p)-\Xi^*(-p)}{2}    & -i\omega_n-\xi_\bp e^{-i\omega_n0^+}
              \end{bmatrix}
\label{invG} 
\end{equation}
where $0^+$ is an infinitesimal positive number. The inverse matrix $V_{\rm ind}^{-1}(p-p')$ is defined by  $\sum_q V_{\rm ind}^{-1}(p-q)V_{\rm ind}(q-p') = \delta_{p p'}$. 
In Eq.\ (\ref{Seff}), $\text{Tr}$ denotes the trace over the $2\times2$
matrix  $\ln G^{-1}(p)$ and the summation over $p$. The mean-field theory for the pairing is obtained from the stationary phase condition
$ \left.{\delta}S_{\rm eff}(\Xi^*,\Xi)/{\delta \Xi^*(p)}\right|_ { \Xi_s} = 0$. Defining the gap as $\Delta(p) = \left[{ \Xi_s(p)- \Xi_s(-p)}\right ]/{2} $, we 
obtain
\begin{align}
\Delta(\bp,i\omega_n) &= -T \sum_{m} \int\frac{d\bq}{(2\pi)^2}V_{\rm ind}(\bp-\bq,i\omega_n-i\omega_m)  \nn \\
&\qquad\times\frac{\Delta(\bq,i\omega_m) }{\omega_m^2+\calE^2(\bq,i\omega_m)}
\label{gapeq_fre}
\end{align}
and
\begin{align}
n_F = {T}\sum_{n}\int\frac{d\bq}{(2\pi)^2} \frac{-i\omega_n e^{i\omega_n 0^+}-\xi_\bq}{\omega_n^2+\calE^2(\bq,i\omega_n)},
\label{par_den}
\end{align}
where $\calE(\bq,i\omega_m) = \sqrt{\xi_\bq^2 + |\Delta(\bq,i\omega_m)|^2}$. Equations (\ref{gapeq_fre}) and (\ref{par_den}) retain the full frequency dependence of the gap and thus constitute a type of Eliashberg theory~\cite{AGD}. When the frequency dependence of the 
induced interaction and the gap is ignored, Eq.\ (\ref{gapeq_fre}) simplifies to the standard BCS gap equation. However, we shall see that retardation effects are generally important.  
 
The pairing between the spin-polarised fermions has  $p_x\pm ip_y$ symmetry due to the rotational symmetry of the system~\cite{Anderson1961}. We thus look for a solution to the gap equation of the form $ \Delta(\bp,i\omega_n) = \Delta_1(p,i\omega_n)e^{i\phi_\bp}$, 
  where $\phi_\bp$ is the azimuthal angle of $\bp$. It follows that only the $p$-wave component of the induced interaction is relevant, which is defined as
 \beq
 V_1(p,q;i\omega_\nu) = \int_0^{2\pi}\frac{d\varphi}{2\pi}V_{\rm ind}(\bp-\bq;i\omega_n-i\omega_m) e^{-i\varphi},
 \label{V1}
 \eeq
 where $\varphi \equiv \phi_\bp-\phi_\bq$. Substituting the $p$-wave form of the gap parameter into the Eq.~(\ref{gapeq_fre}) one easily obtains an equation involving 
 $V_1(p,q;i\omega_\nu)$ which determines the amplitude $\Delta_1(p,i\omega_n)$. Crucial to the feasibility of our numerical solution to this gap equation, we are able to obtain an 
 \emph{analytic} expression for $V_1(p,q;i\omega_\nu)$ in terms of the complete elliptic integrals of the first and second kind. However, this  expression is  complex and will be provided in the Supplementary Material.  
 
\begin{center}
\begin{figure}[ht]
\includegraphics[width=0.8\linewidth]{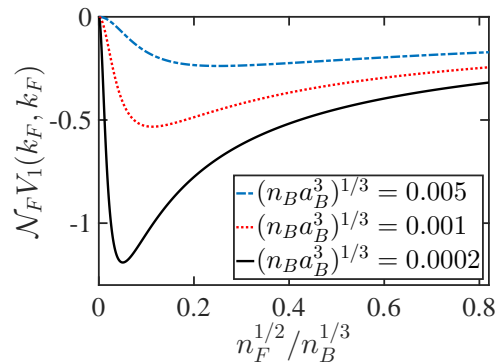}
\caption{The zero-frequency $p$-wave interaction at the Fermi surface  as a function of $n_F^{1/2}/n_B^{1/3}$ for a weakly interacting $^{40} \rm K$-$^7 \rm Li$ mixture with 
$(n_Ba_{\rm eff}^3)^{1/3}=0.1$.  Here  $\calN_F = m_F/2\pi$ is the density of states of the Fermi gas.}
\label{NV}
\end{figure}
\end{center}

A  powerful attribute of the present system is that both the \emph{strength} and the \emph{range} of the induced interaction can be controlled experimentally. 
From Eq.~(\ref{Vinddef}), we see that the induced interaction is proportional to the second power of  $(n_Ba_{\rm eff}^3)^{1/3}$, which measures the strength of the Fermi-Bose interaction. This parameter will be kept small compared to unity corresponding to a weak Fermi-Bose interaction.  In addition to $(n_Ba_{\rm eff}^3)^{1/3}$, the induced interaction depends on the following three dimensionless, independent 
parameters: the Bose gas parameter $(n_Ba_B^3)^{1/3}$,  the ratio of the inter-particle distances
$n_F^{1/2}/n_B^{1/3}$, and finally the mass ratio $m_F/m_B$. To gain some intuition, we plot in Fig.~\ref{NV} the zero-frequency component of the $p$-wave interaction at the Fermi surface, $V_1(k_F,k_F)$, as a function of $n_F^{1/2}/n_B^{1/3}$
   for a mixture of $^{40}{\rm K}$ and $^7{\rm Li}$.  We see that the magnitude of
 $ V_1(k_F,k_F)$ generally exhibits a maximum at a certain value of $n_F^{1/2}/n_B^{1/3}$ for fixed values of  $(n_Ba_{\rm eff}^3)^{1/3}$ and $(n_Ba_{B}^3)^{1/3}$. This non-monotonic behaviour can be understood as follows. 
 In the limit $n_F^{1/2}/n_B^{1/3}\rightarrow\infty$, the density of the Bose gas vanishes compared to that of the Fermi gas, and the induced interaction vanishes as a result. On the other hand, as  $n_F^{1/2}/n_B^{1/3}\rightarrow 0$, the density of the Bose gas increases and the BEC coherence length $\xi_B$ decreases (albeit keeping the Bose gas parameter constant). Since $\xi_B$ determines the range of the induced interaction, the latter eventually becomes short ranged when $\xi_B$ becomes small compared to distances between the fermions. This in turn leads to a suppressed $p$-wave interaction as we see in Fig.~\ref{NV}. Now, the same argument also suggests that the magnitude of $V_1(k_F,k_F)$ monotonically decreases as $(n_Ba_B^3)^{1/3}$ increases for fixed values of $n_F^{1/2}/n_B^{1/3}$, since such a variation leads to a steady decrease of the BEC coherence length.  This in fact is consistent with what  we observe in Fig.~\ref{NV}. 

\begin{center}
\begin{figure}[ht]
\includegraphics[width=0.8\linewidth]{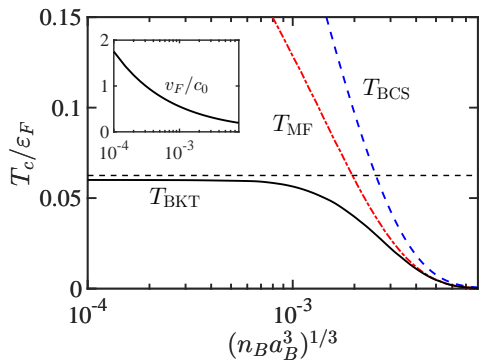}
\caption{The superfluid critical temperatures as a function of $(n_Ba_B^3)^{1/3}$ (on a logarithmic scale) for the  $^{40} \rm K$-$^7 \rm Li$ mixture 
with $n_F^{1/2}/n_B^{1/3}=0.1$ and $(n_Ba_{\rm eff}^3)^{1/3}=0.1$. The inset shows the behaviour of $v_F/c_0$ as a function of $(n_Ba_B^3)^{1/3}$. }
\label{Tc}
\end{figure}
\end{center}

With all the frequency components of the $p$-wave induced interaction $V_1(p,q;i\omega_\nu)$ determined, the mean-field superfluid transition temperature 
$ T_{\rm MF}$ can be obtained by solving the linearised gap equation. As an example, $T_{\rm MF}$ is shown in Fig.~\ref{Tc} 
for a mixture of $^{40}{\rm K}$ and $^7{\rm Li}$ as a function of
the BEC gas parameter $(n_Ba_B^3)^{1/3}$  for $n_F^{1/2}/n_B^{1/3}=0.1$ and a weak Fermi-Bose interaction strength  $(n_Ba_{\rm eff}^3)^{1/3}=0.1$. 
We see that the transition temperature increases with a decreasing $(n_Ba_B^3)^{1/3}$ in agreement with the previous analysis concerning
 the strength of $p$-wave interaction. In fact, the mean-field 
transition temperature becomes significant compared to the Fermi energy for small $(n_Ba_B^3)^{1/3}$.
This is a promising result, even though phase fluctuations will reduce the critical temperature significantly as we shall see shortly. 
We also show the critical temperature $T_{\rm BCS}$ obtained from the BCS theory by neglecting the frequency dependence of the induced 
 interaction, i.e., by using the $i\omega_\nu=0$ component of the induced interaction in the gap equation.  We see that these two temperatures indeed agree when $v_F/c_0\ll 1$,
  as we have argued previously, whereas retardation effects significantly suppress the pairing for larger $v_F/c_0$. 

\emph{BKT transition temperature.--}   
Since the Fermi system is 2D, phase fluctuations of the order parameter can significantly suppress the critical  temperature of the 
superfluid transition. The transition is  described by BKT theory, where  the critical temperature is determined by the condition~\cite{Berezinskii1972,Kosterlitz1973,Kosterlitz1974,Jose1977}
\begin{align}
T_{\rm BKT} = \frac{\pi}{8m_F^2}\rho_s\left (\left \{\Delta(i\omega_n)\right \}, T_{\rm BKT}\right ).
\label{TBKT}
\end{align}
Here  $\rho_s$ is the superfluid mass density, which can be defined by considering the free energy density $F=\Omega + \mu n_F$ of the superfluid flowing at a velocity $\mathbf{v}$, where $\Omega=-T\ln{\mathcal Z}/{\mathcal A}$ is the grand potential density. For small velocities, we have $F(\mathbf{v})-F(0)=\rho_sv^2/2$. Using $(\partial_v^2F)_{n_F}=(\partial_v^2\Omega)_{\mu}$~\cite{Taylor2006} and $\rho_s=\left.\partial_v^2F({\bv})\right|_{\bv=0}$~\cite{Fisher1973,Lieb2002}, the superfluid density can be obtained as $\rho_s=\left.\partial_v^2\Omega({\bv})\right|_{\bv=0}$.  Within mean field theory, we have  $\Omega_\text{MF} (\bv)=TS_\text{eff} (\bv)/{\mathcal A}$, where $S_\text{eff} (\bv)$ can be obtained from Eq.~(\ref{Seff}) by means of  a momentum boost of ${\mathbf q}=m_F\bv$. The momentum boost enters only in the diagonal components of $G^{-1}(p)$ in Eq.~(\ref{invG}), which now read $-i\omega_n+\xi_{\bp+\bq} e^{i\omega_n0^+}$ and $-i\omega_n-\xi_{\bp-\bq} e^{-i\omega_n0^+}$. A straightforward evaluation of $\partial_v^2\Omega_{\rm MF}(\bv)|_{\bv=0}$ yields (see Supplementary Material)
\begin{align}
\rho_s = \rho_0 + \frac{T}{2}\sum_{n}\int \frac{d\bp}{(2\pi)^2} p^2 \frac{ \calE^2(\bp,i\omega_n)-\omega_n^2}{\left[\omega_n^2+\calE^2(\bp,i\omega_n)\right ]^2},
\label{ns}
\end{align} 
where $\rho_0 = m_F n_F$. In the case of frequency-independent gap parameters $\Delta_\bp$, Eq.~(\ref{ns}) reduces to the well-known result
$\rho_s = \rho_0 - (8T)^{-1}\int d\bp(2\pi)^{-2} p^2 {\rm sech}^2( \calE_\bp/2T)$.

\begin{center}
\begin{figure}[ht]
\includegraphics[width=0.8\linewidth]{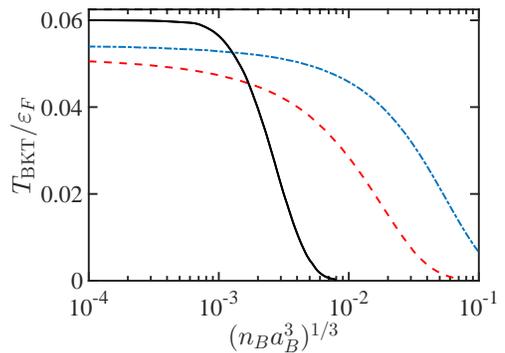}
\caption{The BKT temperatures as a function of $(n_Ba_B^3)^{1/3}$ (on a logarithmic scale) for the  $^{40} \rm K$-$^7 \rm Li$ mixture for various parameter settings, namely for $(a)$ $n_F^{1/2}/n_B^{1/3}=0.1$ and $(n_Ba_{\rm eff}^3)^{1/3}=0.1$ (black solid), $(b)$ $n_F^{1/2}/n_B^{1/3}=0.5$ and $(n_Ba_{\rm eff}^3)^{1/3}=0.15$ (red dashed), and $(c)$ $n_F^{1/2}/n_B^{1/3}=0.8$ and $(n_Ba_{\rm eff}^3)^{1/3}=0.2$ (blue dash-dot).}
\label{Tc2}
\end{figure}
\end{center}

The BKT transition temperature can now be obtained by solving Eqs.~(\ref{gapeq_fre}), (\ref{TBKT}) and (\ref{ns}) self-consistently~\cite{footnote2}.
The superfluid mass density $\rho_s$ equals the total Fermi mass density $\rho_0$ at $T = 0$ and gradually decreases when $T$ increases. 
We find that for a small Bose gas parameter, which corresponds to a long range induced interaction, 
the reduction in $\rho_s(T)$ by increasing temperature is rather small and $\rho_s(T)\simeq \rho_0$ when the BKT melting condition Eq.\ (\ref{TBKT}) is fulfilled. 
This suggests that the $T_{\rm BKT} $ in such a scenario will be close to the \emph{maximum value allowed by BKT theory}, i.e., $T_{\rm BKT}= \varepsilon_F/16$. We point out that 
this value is in fact within experimental reach~\cite{footnote}. In Fig.~\ref{Tc}, 
the BKT transition temperature is shown as a function of $(n_Ba_B^3)^{1/3}$ for the previously given physical parameters. The transition temperature indeed quickly reaches the limiting value 
$\varepsilon_F/16$ as  $(n_Ba_B^3)^{1/3}$ decreases. Importantly, this maximum value is reached for a  weak Fermi-Bose coupling with $(n_Ba_{\rm eff}^3)^{1/3}=0.1$, 
which means that possible three-body losses due to a dimer formation from a fermion and a boson are small. Finally we emphasise that our system is very flexible in the sense that 
 high transition temperatures can be reached across a \emph{broad range} of parameter space. This 
is shown in Fig.~\ref{Tc2}, where BKT temperatures for various settings of parameters, all within the weak Fermi-Bose interaction regime, are calculated. We see that transition temperatures in the vicinity of the BKT limiting value are achieved for various density ratios of the mixture. Such a high flexibility 
makes our system advantageous in comparison to other proposals to realise $p$-wave superfluid in cold atomic systems. 

\emph{Conclusions.--}  We have shown that a 2D-3D Bose-Fermi mixture is a  promising
  system to  realise a topological $p_x+ip_y$ superfluid. The fermions attract each other via density modulations in the BEC and form $p$-wave paring. We analyse the paring by solving the frequency-dependent gap equation which takes the retardation effects fully into account. The resulting Eliashberg theory was then combined with BKT theory to obtain a reliable microscopic theory for the superfluid critical temperature. 
 Both the strength and the range of the induced interaction  between the fermions  can be controlled, and this can be used to tune the critical temperature 
 to a limiting value imposed by BKT theory. Importantly, this maximum can be reached while keeping the Bose-Fermi interaction weak. Our results are directly relevant for experiments which use cold atomic gases to explore the topological superfluids.  

\begin{acknowledgments}
 G.M.B.~wishes
  to acknowledge the support of the Villum Foundation via grant
  VKR023163. We  acknowledge helpful discussions with Ed Taylor and Jonatan Midtgaard. 
\end{acknowledgments}

\bibliographystyle{apsrev4-1}

\begin{thebibliography}{45}%
\makeatletter
\providecommand \@ifxundefined [1]{%
 \@ifx{#1\undefined}
}%
\providecommand \@ifnum [1]{%
 \ifnum #1\expandafter \@firstoftwo
 \else \expandafter \@secondoftwo
 \fi
}%
\providecommand \@ifx [1]{%
 \ifx #1\expandafter \@firstoftwo
 \else \expandafter \@secondoftwo
 \fi
}%
\providecommand \natexlab [1]{#1}%
\providecommand \enquote  [1]{``#1''}%
\providecommand \bibnamefont  [1]{#1}%
\providecommand \bibfnamefont [1]{#1}%
\providecommand \citenamefont [1]{#1}%
\providecommand \href@noop [0]{\@secondoftwo}%
\providecommand \href [0]{\begingroup \@sanitize@url \@href}%
\providecommand \@href[1]{\@@startlink{#1}\@@href}%
\providecommand \@@href[1]{\endgroup#1\@@endlink}%
\providecommand \@sanitize@url [0]{\catcode `\\12\catcode `\$12\catcode
  `\&12\catcode `\#12\catcode `\^12\catcode `\_12\catcode `\%12\relax}%
\providecommand \@@startlink[1]{}%
\providecommand \@@endlink[0]{}%
\providecommand \url  [0]{\begingroup\@sanitize@url \@url }%
\providecommand \@url [1]{\endgroup\@href {#1}{\urlprefix }}%
\providecommand \urlprefix  [0]{URL }%
\providecommand \Eprint [0]{\href }%
\providecommand \doibase [0]{http://dx.doi.org/}%
\providecommand \selectlanguage [0]{\@gobble}%
\providecommand \bibinfo  [0]{\@secondoftwo}%
\providecommand \bibfield  [0]{\@secondoftwo}%
\providecommand \translation [1]{[#1]}%
\providecommand \BibitemOpen [0]{}%
\providecommand \bibitemStop [0]{}%
\providecommand \bibitemNoStop [0]{.\EOS\space}%
\providecommand \EOS [0]{\spacefactor3000\relax}%
\providecommand \BibitemShut  [1]{\csname bibitem#1\endcsname}%
\let\auto@bib@innerbib\@empty
\bibitem [{\citenamefont {Hasan}\ and\ \citenamefont {Kane}(2010)}]{Hasan2010}%
  \BibitemOpen
  \bibfield  {author} {\bibinfo {author} {\bibfnamefont {M.~Z.}\ \bibnamefont
  {Hasan}}\ and\ \bibinfo {author} {\bibfnamefont {C.~L.}\ \bibnamefont
  {Kane}},\ }\href {\doibase 10.1103/RevModPhys.82.3045} {\bibfield  {journal}
  {\bibinfo  {journal} {Rev. Mod. Phys.}\ }\textbf {\bibinfo {volume} {82}},\
  \bibinfo {pages} {3045} (\bibinfo {year} {2010})}\BibitemShut {NoStop}%
\bibitem [{\citenamefont {Qi}\ and\ \citenamefont {Zhang}(2011)}]{Qi2011}%
  \BibitemOpen
  \bibfield  {author} {\bibinfo {author} {\bibfnamefont {X.-L.}\ \bibnamefont
  {Qi}}\ and\ \bibinfo {author} {\bibfnamefont {S.-C.}\ \bibnamefont {Zhang}},\
  }\href {\doibase 10.1103/RevModPhys.83.1057} {\bibfield  {journal} {\bibinfo
  {journal} {Rev. Mod. Phys.}\ }\textbf {\bibinfo {volume} {83}},\ \bibinfo
  {pages} {1057} (\bibinfo {year} {2011})}\BibitemShut {NoStop}%
\bibitem [{\citenamefont {Nayak}\ \emph {et~al.}(2008)\citenamefont {Nayak},
  \citenamefont {Simon}, \citenamefont {Stern}, \citenamefont {Freedman},\ and\
  \citenamefont {Das~Sarma}}]{Nayak2008}%
  \BibitemOpen
  \bibfield  {author} {\bibinfo {author} {\bibfnamefont {C.}~\bibnamefont
  {Nayak}}, \bibinfo {author} {\bibfnamefont {S.~H.}\ \bibnamefont {Simon}},
  \bibinfo {author} {\bibfnamefont {A.}~\bibnamefont {Stern}}, \bibinfo
  {author} {\bibfnamefont {M.}~\bibnamefont {Freedman}}, \ and\ \bibinfo
  {author} {\bibfnamefont {S.}~\bibnamefont {Das~Sarma}},\ }\href {\doibase
  10.1103/RevModPhys.80.1083} {\bibfield  {journal} {\bibinfo  {journal} {Rev.
  Mod. Phys.}\ }\textbf {\bibinfo {volume} {80}},\ \bibinfo {pages} {1083}
  (\bibinfo {year} {2008})}\BibitemShut {NoStop}%
\bibitem [{\citenamefont {Mourik}\ \emph {et~al.}(2012)\citenamefont {Mourik},
  \citenamefont {Zuo}, \citenamefont {Frolov}, \citenamefont {Plissard},
  \citenamefont {Bakkers},\ and\ \citenamefont {Kouwenhoven}}]{Mourik2012}%
  \BibitemOpen
  \bibfield  {author} {\bibinfo {author} {\bibfnamefont {V.}~\bibnamefont
  {Mourik}}, \bibinfo {author} {\bibfnamefont {K.}~\bibnamefont {Zuo}},
  \bibinfo {author} {\bibfnamefont {S.~M.}\ \bibnamefont {Frolov}}, \bibinfo
  {author} {\bibfnamefont {S.~R.}\ \bibnamefont {Plissard}}, \bibinfo {author}
  {\bibfnamefont {E.~P. A.~M.}\ \bibnamefont {Bakkers}}, \ and\ \bibinfo
  {author} {\bibfnamefont {L.~P.}\ \bibnamefont {Kouwenhoven}},\ }\href
  {\doibase 10.1126/science.1222360} {\bibfield  {journal} {\bibinfo  {journal}
  {Science}\ }\textbf {\bibinfo {volume} {336}},\ \bibinfo {pages} {1003}
  (\bibinfo {year} {2012})},\ \Eprint
  {http://arxiv.org/abs/http://science.sciencemag.org/content/336/6084/1003.full.pdf}
  {http://science.sciencemag.org/content/336/6084/1003.full.pdf} \BibitemShut
  {NoStop}%
\bibitem [{\citenamefont {Deng}\ \emph {et~al.}(2012)\citenamefont {Deng},
  \citenamefont {Yu}, \citenamefont {Huang}, \citenamefont {Larsson},
  \citenamefont {Caroff},\ and\ \citenamefont {Xu}}]{Deng2012}%
  \BibitemOpen
  \bibfield  {author} {\bibinfo {author} {\bibfnamefont {M.~T.}\ \bibnamefont
  {Deng}}, \bibinfo {author} {\bibfnamefont {C.~L.}\ \bibnamefont {Yu}},
  \bibinfo {author} {\bibfnamefont {G.~Y.}\ \bibnamefont {Huang}}, \bibinfo
  {author} {\bibfnamefont {M.}~\bibnamefont {Larsson}}, \bibinfo {author}
  {\bibfnamefont {P.}~\bibnamefont {Caroff}}, \ and\ \bibinfo {author}
  {\bibfnamefont {H.~Q.}\ \bibnamefont {Xu}},\ }\href {\doibase
  10.1021/nl303758w} {\bibfield  {journal} {\bibinfo  {journal} {Nano Letters}\
  }\textbf {\bibinfo {volume} {12}},\ \bibinfo {pages} {6414} (\bibinfo {year}
  {2012})},\ \bibinfo {note} {pMID: 23181691},\ \Eprint
  {http://arxiv.org/abs/http://dx.doi.org/10.1021/nl303758w}
  {http://dx.doi.org/10.1021/nl303758w} \BibitemShut {NoStop}%
\bibitem [{\citenamefont {Das}\ \emph {et~al.}(2012)\citenamefont {Das},
  \citenamefont {Ronen}, \citenamefont {Most}, \citenamefont {Oreg},
  \citenamefont {Heiblum},\ and\ \citenamefont {Shtrikman}}]{Das2012}%
  \BibitemOpen
  \bibfield  {author} {\bibinfo {author} {\bibfnamefont {A.}~\bibnamefont
  {Das}}, \bibinfo {author} {\bibfnamefont {Y.}~\bibnamefont {Ronen}}, \bibinfo
  {author} {\bibfnamefont {Y.}~\bibnamefont {Most}}, \bibinfo {author}
  {\bibfnamefont {Y.}~\bibnamefont {Oreg}}, \bibinfo {author} {\bibfnamefont
  {M.}~\bibnamefont {Heiblum}}, \ and\ \bibinfo {author} {\bibfnamefont
  {H.}~\bibnamefont {Shtrikman}},\ }\href {http://dx.doi.org/10.1038/nphys2479}
  {\bibfield  {journal} {\bibinfo  {journal} {Nat Phys}\ }\textbf {\bibinfo
  {volume} {8}},\ \bibinfo {pages} {887} (\bibinfo {year} {2012})}\BibitemShut
  {NoStop}%
\bibitem [{\citenamefont {Rokhinson}\ \emph {et~al.}(2012)\citenamefont
  {Rokhinson}, \citenamefont {Liu},\ and\ \citenamefont
  {Furdyna}}]{Rokhinson2012}%
  \BibitemOpen
  \bibfield  {author} {\bibinfo {author} {\bibfnamefont {L.~P.}\ \bibnamefont
  {Rokhinson}}, \bibinfo {author} {\bibfnamefont {X.}~\bibnamefont {Liu}}, \
  and\ \bibinfo {author} {\bibfnamefont {J.~K.}\ \bibnamefont {Furdyna}},\
  }\href {http://dx.doi.org/10.1038/nphys2429} {\bibfield  {journal} {\bibinfo
  {journal} {Nat Phys}\ }\textbf {\bibinfo {volume} {8}},\ \bibinfo {pages}
  {795} (\bibinfo {year} {2012})}\BibitemShut {NoStop}%
\bibitem [{\citenamefont {Finck}\ \emph {et~al.}(2013)\citenamefont {Finck},
  \citenamefont {Van~Harlingen}, \citenamefont {Mohseni}, \citenamefont
  {Jung},\ and\ \citenamefont {Li}}]{Finck2013}%
  \BibitemOpen
  \bibfield  {author} {\bibinfo {author} {\bibfnamefont {A.~D.~K.}\
  \bibnamefont {Finck}}, \bibinfo {author} {\bibfnamefont {D.~J.}\ \bibnamefont
  {Van~Harlingen}}, \bibinfo {author} {\bibfnamefont {P.~K.}\ \bibnamefont
  {Mohseni}}, \bibinfo {author} {\bibfnamefont {K.}~\bibnamefont {Jung}}, \
  and\ \bibinfo {author} {\bibfnamefont {X.}~\bibnamefont {Li}},\ }\href
  {\doibase 10.1103/PhysRevLett.110.126406} {\bibfield  {journal} {\bibinfo
  {journal} {Phys. Rev. Lett.}\ }\textbf {\bibinfo {volume} {110}},\ \bibinfo
  {pages} {126406} (\bibinfo {year} {2013})}\BibitemShut {NoStop}%
\bibitem [{\citenamefont {Nadj-Perge}\ \emph {et~al.}(2014)\citenamefont
  {Nadj-Perge}, \citenamefont {Drozdov}, \citenamefont {Li}, \citenamefont
  {Chen}, \citenamefont {Jeon}, \citenamefont {Seo}, \citenamefont {MacDonald},
  \citenamefont {Bernevig},\ and\ \citenamefont {Yazdani}}]{NadjPerge2014}%
  \BibitemOpen
  \bibfield  {author} {\bibinfo {author} {\bibfnamefont {S.}~\bibnamefont
  {Nadj-Perge}}, \bibinfo {author} {\bibfnamefont {I.~K.}\ \bibnamefont
  {Drozdov}}, \bibinfo {author} {\bibfnamefont {J.}~\bibnamefont {Li}},
  \bibinfo {author} {\bibfnamefont {H.}~\bibnamefont {Chen}}, \bibinfo {author}
  {\bibfnamefont {S.}~\bibnamefont {Jeon}}, \bibinfo {author} {\bibfnamefont
  {J.}~\bibnamefont {Seo}}, \bibinfo {author} {\bibfnamefont {A.~H.}\
  \bibnamefont {MacDonald}}, \bibinfo {author} {\bibfnamefont {B.~A.}\
  \bibnamefont {Bernevig}}, \ and\ \bibinfo {author} {\bibfnamefont
  {A.}~\bibnamefont {Yazdani}},\ }\href {\doibase 10.1126/science.1259327}
  {\bibfield  {journal} {\bibinfo  {journal} {Science}\ }\textbf {\bibinfo
  {volume} {346}},\ \bibinfo {pages} {602} (\bibinfo {year} {2014})},\ \Eprint
  {http://arxiv.org/abs/http://science.sciencemag.org/content/346/6209/602.full.pdf}
  {http://science.sciencemag.org/content/346/6209/602.full.pdf} \BibitemShut
  {NoStop}%
\bibitem [{\citenamefont {Albrecht}\ \emph {et~al.}(2016)\citenamefont
  {Albrecht}, \citenamefont {Higginbotham}, \citenamefont {Madsen},
  \citenamefont {Kuemmeth}, \citenamefont {Jespersen}, \citenamefont
  {Nyg{\aa}rd}, \citenamefont {Krogstrup},\ and\ \citenamefont
  {Marcus}}]{Albrecht2016}%
  \BibitemOpen
  \bibfield  {author} {\bibinfo {author} {\bibfnamefont {S.~M.}\ \bibnamefont
  {Albrecht}}, \bibinfo {author} {\bibfnamefont {A.~P.}\ \bibnamefont
  {Higginbotham}}, \bibinfo {author} {\bibfnamefont {M.}~\bibnamefont
  {Madsen}}, \bibinfo {author} {\bibfnamefont {F.}~\bibnamefont {Kuemmeth}},
  \bibinfo {author} {\bibfnamefont {T.~S.}\ \bibnamefont {Jespersen}}, \bibinfo
  {author} {\bibfnamefont {J.}~\bibnamefont {Nyg{\aa}rd}}, \bibinfo {author}
  {\bibfnamefont {P.}~\bibnamefont {Krogstrup}}, \ and\ \bibinfo {author}
  {\bibfnamefont {C.~M.}\ \bibnamefont {Marcus}},\ }\href
  {http://dx.doi.org/10.1038/nature17162} {\bibfield  {journal} {\bibinfo
  {journal} {Nature}\ }\textbf {\bibinfo {volume} {531}},\ \bibinfo {pages}
  {206} (\bibinfo {year} {2016})}\BibitemShut {NoStop}%
\bibitem [{\citenamefont {Ishida}\ \emph {et~al.}(1998)\citenamefont {Ishida},
  \citenamefont {Mukuda}, \citenamefont {Kitaoka}, \citenamefont {Asayama},
  \citenamefont {Mao}, \citenamefont {Mori},\ and\ \citenamefont
  {Maeno}}]{Ishida1998}%
  \BibitemOpen
  \bibfield  {author} {\bibinfo {author} {\bibfnamefont {K.}~\bibnamefont
  {Ishida}}, \bibinfo {author} {\bibfnamefont {H.}~\bibnamefont {Mukuda}},
  \bibinfo {author} {\bibfnamefont {Y.}~\bibnamefont {Kitaoka}}, \bibinfo
  {author} {\bibfnamefont {K.}~\bibnamefont {Asayama}}, \bibinfo {author}
  {\bibfnamefont {Z.~Q.}\ \bibnamefont {Mao}}, \bibinfo {author} {\bibfnamefont
  {Y.}~\bibnamefont {Mori}}, \ and\ \bibinfo {author} {\bibfnamefont
  {Y.}~\bibnamefont {Maeno}},\ }\href {http://dx.doi.org/10.1038/25315}
  {\bibfield  {journal} {\bibinfo  {journal} {Nature}\ }\textbf {\bibinfo
  {volume} {396}},\ \bibinfo {pages} {658} (\bibinfo {year}
  {1998})}\BibitemShut {NoStop}%
\bibitem [{\citenamefont {Nelson}\ \emph {et~al.}(2004)\citenamefont {Nelson},
  \citenamefont {Mao}, \citenamefont {Maeno},\ and\ \citenamefont
  {Liu}}]{Nelson2004}%
  \BibitemOpen
  \bibfield  {author} {\bibinfo {author} {\bibfnamefont {K.~D.}\ \bibnamefont
  {Nelson}}, \bibinfo {author} {\bibfnamefont {Z.~Q.}\ \bibnamefont {Mao}},
  \bibinfo {author} {\bibfnamefont {Y.}~\bibnamefont {Maeno}}, \ and\ \bibinfo
  {author} {\bibfnamefont {Y.}~\bibnamefont {Liu}},\ }\href {\doibase
  10.1126/science.1103881} {\bibfield  {journal} {\bibinfo  {journal}
  {Science}\ }\textbf {\bibinfo {volume} {306}},\ \bibinfo {pages} {1151}
  (\bibinfo {year} {2004})},\ \Eprint
  {http://arxiv.org/abs/http://science.sciencemag.org/content/306/5699/1151.full.pdf}
  {http://science.sciencemag.org/content/306/5699/1151.full.pdf} \BibitemShut
  {NoStop}%
\bibitem [{\citenamefont {Kidwingira}\ \emph {et~al.}(2006)\citenamefont
  {Kidwingira}, \citenamefont {Strand}, \citenamefont {Van~Harlingen},\ and\
  \citenamefont {Maeno}}]{Kidwingira2006}%
  \BibitemOpen
  \bibfield  {author} {\bibinfo {author} {\bibfnamefont {F.}~\bibnamefont
  {Kidwingira}}, \bibinfo {author} {\bibfnamefont {J.~D.}\ \bibnamefont
  {Strand}}, \bibinfo {author} {\bibfnamefont {D.~J.}\ \bibnamefont
  {Van~Harlingen}}, \ and\ \bibinfo {author} {\bibfnamefont {Y.}~\bibnamefont
  {Maeno}},\ }\href {\doibase 10.1126/science.1133239} {\bibfield  {journal}
  {\bibinfo  {journal} {Science}\ }\textbf {\bibinfo {volume} {314}},\ \bibinfo
  {pages} {1267} (\bibinfo {year} {2006})},\ \Eprint
  {http://arxiv.org/abs/http://science.sciencemag.org/content/314/5803/1267.full.pdf}
  {http://science.sciencemag.org/content/314/5803/1267.full.pdf} \BibitemShut
  {NoStop}%
\bibitem [{\citenamefont {Xia}\ \emph {et~al.}(2006)\citenamefont {Xia},
  \citenamefont {Maeno}, \citenamefont {Beyersdorf}, \citenamefont {Fejer},\
  and\ \citenamefont {Kapitulnik}}]{Xia2006}%
  \BibitemOpen
  \bibfield  {author} {\bibinfo {author} {\bibfnamefont {J.}~\bibnamefont
  {Xia}}, \bibinfo {author} {\bibfnamefont {Y.}~\bibnamefont {Maeno}}, \bibinfo
  {author} {\bibfnamefont {P.~T.}\ \bibnamefont {Beyersdorf}}, \bibinfo
  {author} {\bibfnamefont {M.~M.}\ \bibnamefont {Fejer}}, \ and\ \bibinfo
  {author} {\bibfnamefont {A.}~\bibnamefont {Kapitulnik}},\ }\href {\doibase
  10.1103/PhysRevLett.97.167002} {\bibfield  {journal} {\bibinfo  {journal}
  {Phys. Rev. Lett.}\ }\textbf {\bibinfo {volume} {97}},\ \bibinfo {pages}
  {167002} (\bibinfo {year} {2006})}\BibitemShut {NoStop}%
\bibitem [{\citenamefont {Gurarie}\ and\ \citenamefont
  {Radzihovsky}(2007)}]{Gurarie2007}%
  \BibitemOpen
  \bibfield  {author} {\bibinfo {author} {\bibfnamefont {V.}~\bibnamefont
  {Gurarie}}\ and\ \bibinfo {author} {\bibfnamefont {L.}~\bibnamefont
  {Radzihovsky}},\ }\href {\doibase
  http://dx.doi.org/10.1016/j.aop.2006.10.009} {\bibfield  {journal} {\bibinfo
  {journal} {Annals of Physics}\ }\textbf {\bibinfo {volume} {322}},\ \bibinfo
  {pages} {2 } (\bibinfo {year} {2007})},\ \bibinfo {note} {january Special
  Issue 2007}\BibitemShut {NoStop}%
\bibitem [{\citenamefont {G\"unter}\ \emph {et~al.}(2005)\citenamefont
  {G\"unter}, \citenamefont {St\"oferle}, \citenamefont {Moritz}, \citenamefont
  {K\"ohl},\ and\ \citenamefont {Esslinger}}]{Gunter2005}%
  \BibitemOpen
  \bibfield  {author} {\bibinfo {author} {\bibfnamefont {K.}~\bibnamefont
  {G\"unter}}, \bibinfo {author} {\bibfnamefont {T.}~\bibnamefont
  {St\"oferle}}, \bibinfo {author} {\bibfnamefont {H.}~\bibnamefont {Moritz}},
  \bibinfo {author} {\bibfnamefont {M.}~\bibnamefont {K\"ohl}}, \ and\ \bibinfo
  {author} {\bibfnamefont {T.}~\bibnamefont {Esslinger}},\ }\href {\doibase
  10.1103/PhysRevLett.95.230401} {\bibfield  {journal} {\bibinfo  {journal}
  {Phys. Rev. Lett.}\ }\textbf {\bibinfo {volume} {95}},\ \bibinfo {pages}
  {230401} (\bibinfo {year} {2005})}\BibitemShut {NoStop}%
\bibitem [{\citenamefont {Gaebler}\ \emph {et~al.}(2007)\citenamefont
  {Gaebler}, \citenamefont {Stewart}, \citenamefont {Bohn},\ and\ \citenamefont
  {Jin}}]{Gaebler2007}%
  \BibitemOpen
  \bibfield  {author} {\bibinfo {author} {\bibfnamefont {J.~P.}\ \bibnamefont
  {Gaebler}}, \bibinfo {author} {\bibfnamefont {J.~T.}\ \bibnamefont
  {Stewart}}, \bibinfo {author} {\bibfnamefont {J.~L.}\ \bibnamefont {Bohn}}, \
  and\ \bibinfo {author} {\bibfnamefont {D.~S.}\ \bibnamefont {Jin}},\ }\href
  {\doibase 10.1103/PhysRevLett.98.200403} {\bibfield  {journal} {\bibinfo
  {journal} {Phys. Rev. Lett.}\ }\textbf {\bibinfo {volume} {98}},\ \bibinfo
  {pages} {200403} (\bibinfo {year} {2007})}\BibitemShut {NoStop}%
\bibitem [{\citenamefont {Levinsen}\ \emph {et~al.}(2007)\citenamefont
  {Levinsen}, \citenamefont {Cooper},\ and\ \citenamefont
  {Gurarie}}]{Levinsen2007}%
  \BibitemOpen
  \bibfield  {author} {\bibinfo {author} {\bibfnamefont {J.}~\bibnamefont
  {Levinsen}}, \bibinfo {author} {\bibfnamefont {N.~R.}\ \bibnamefont
  {Cooper}}, \ and\ \bibinfo {author} {\bibfnamefont {V.}~\bibnamefont
  {Gurarie}},\ }\href {\doibase 10.1103/PhysRevLett.99.210402} {\bibfield
  {journal} {\bibinfo  {journal} {Phys. Rev. Lett.}\ }\textbf {\bibinfo
  {volume} {99}},\ \bibinfo {pages} {210402} (\bibinfo {year}
  {2007})}\BibitemShut {NoStop}%
\bibitem [{\citenamefont {Jona-Lasinio}\ \emph {et~al.}(2008)\citenamefont
  {Jona-Lasinio}, \citenamefont {Pricoupenko},\ and\ \citenamefont
  {Castin}}]{JonaLasinio2008}%
  \BibitemOpen
  \bibfield  {author} {\bibinfo {author} {\bibfnamefont {M.}~\bibnamefont
  {Jona-Lasinio}}, \bibinfo {author} {\bibfnamefont {L.}~\bibnamefont
  {Pricoupenko}}, \ and\ \bibinfo {author} {\bibfnamefont {Y.}~\bibnamefont
  {Castin}},\ }\href {\doibase 10.1103/PhysRevA.77.043611} {\bibfield
  {journal} {\bibinfo  {journal} {Phys. Rev. A}\ }\textbf {\bibinfo {volume}
  {77}},\ \bibinfo {pages} {043611} (\bibinfo {year} {2008})}\BibitemShut
  {NoStop}%
\bibitem [{\citenamefont {B{\"u}hler}\ \emph {et~al.}(2014)\citenamefont
  {B{\"u}hler}, \citenamefont {Lang}, \citenamefont {Kraus}, \citenamefont
  {M{\"o}ller}, \citenamefont {Huber},\ and\ \citenamefont
  {B{\"u}chler}}]{Buhler2014}%
  \BibitemOpen
  \bibfield  {author} {\bibinfo {author} {\bibfnamefont {A.}~\bibnamefont
  {B{\"u}hler}}, \bibinfo {author} {\bibfnamefont {N.}~\bibnamefont {Lang}},
  \bibinfo {author} {\bibfnamefont {C.~V.}\ \bibnamefont {Kraus}}, \bibinfo
  {author} {\bibfnamefont {G.}~\bibnamefont {M{\"o}ller}}, \bibinfo {author}
  {\bibfnamefont {S.~D.}\ \bibnamefont {Huber}}, \ and\ \bibinfo {author}
  {\bibfnamefont {H.~P.}\ \bibnamefont {B{\"u}chler}},\ }\href
  {http://dx.doi.org/10.1038/ncomms5504} {\bibfield  {journal} {\bibinfo
  {journal} {Nat Commun}\ }\textbf {\bibinfo {volume} {5}} (\bibinfo {year}
  {2014})}\BibitemShut {NoStop}%
\bibitem [{\citenamefont {Massignan}\ \emph {et~al.}(2010)\citenamefont
  {Massignan}, \citenamefont {Sanpera},\ and\ \citenamefont
  {Lewenstein}}]{Massignan2010}%
  \BibitemOpen
  \bibfield  {author} {\bibinfo {author} {\bibfnamefont {P.}~\bibnamefont
  {Massignan}}, \bibinfo {author} {\bibfnamefont {A.}~\bibnamefont {Sanpera}},
  \ and\ \bibinfo {author} {\bibfnamefont {M.}~\bibnamefont {Lewenstein}},\
  }\href {\doibase 10.1103/PhysRevA.81.031607} {\bibfield  {journal} {\bibinfo
  {journal} {Phys. Rev. A}\ }\textbf {\bibinfo {volume} {81}},\ \bibinfo
  {pages} {031607} (\bibinfo {year} {2010})}\BibitemShut {NoStop}%
\bibitem [{\citenamefont {Mathey}\ \emph {et~al.}(2006)\citenamefont {Mathey},
  \citenamefont {Tsai},\ and\ \citenamefont {Neto}}]{Mathey2006}%
  \BibitemOpen
  \bibfield  {author} {\bibinfo {author} {\bibfnamefont {L.}~\bibnamefont
  {Mathey}}, \bibinfo {author} {\bibfnamefont {S.-W.}\ \bibnamefont {Tsai}}, \
  and\ \bibinfo {author} {\bibfnamefont {A.~H.~C.}\ \bibnamefont {Neto}},\
  }\href {\doibase 10.1103/PhysRevLett.97.030601} {\bibfield  {journal}
  {\bibinfo  {journal} {Phys. Rev. Lett.}\ }\textbf {\bibinfo {volume} {97}},\
  \bibinfo {pages} {030601} (\bibinfo {year} {2006})}\BibitemShut {NoStop}%
\bibitem [{\citenamefont {Wu}\ \emph {et~al.}(2012)\citenamefont {Wu},
  \citenamefont {He}, \citenamefont {Zang},\ and\ \citenamefont
  {Kou}}]{Wu2012}%
  \BibitemOpen
  \bibfield  {author} {\bibinfo {author} {\bibfnamefont {Y.-J.}\ \bibnamefont
  {Wu}}, \bibinfo {author} {\bibfnamefont {J.}~\bibnamefont {He}}, \bibinfo
  {author} {\bibfnamefont {C.-L.}\ \bibnamefont {Zang}}, \ and\ \bibinfo
  {author} {\bibfnamefont {S.-P.}\ \bibnamefont {Kou}},\ }\href {\doibase
  10.1103/PhysRevB.86.085128} {\bibfield  {journal} {\bibinfo  {journal} {Phys.
  Rev. B}\ }\textbf {\bibinfo {volume} {86}},\ \bibinfo {pages} {085128}
  (\bibinfo {year} {2012})}\BibitemShut {NoStop}%
\bibitem [{\citenamefont {Zhang}\ \emph {et~al.}(2008)\citenamefont {Zhang},
  \citenamefont {Tewari}, \citenamefont {Lutchyn},\ and\ \citenamefont
  {Das~Sarma}}]{Zhang2008}%
  \BibitemOpen
  \bibfield  {author} {\bibinfo {author} {\bibfnamefont {C.}~\bibnamefont
  {Zhang}}, \bibinfo {author} {\bibfnamefont {S.}~\bibnamefont {Tewari}},
  \bibinfo {author} {\bibfnamefont {R.~M.}\ \bibnamefont {Lutchyn}}, \ and\
  \bibinfo {author} {\bibfnamefont {S.}~\bibnamefont {Das~Sarma}},\ }\href
  {\doibase 10.1103/PhysRevLett.101.160401} {\bibfield  {journal} {\bibinfo
  {journal} {Phys. Rev. Lett.}\ }\textbf {\bibinfo {volume} {101}},\ \bibinfo
  {pages} {160401} (\bibinfo {year} {2008})}\BibitemShut {NoStop}%
\bibitem [{\citenamefont {Sato}\ \emph {et~al.}(2009)\citenamefont {Sato},
  \citenamefont {Takahashi},\ and\ \citenamefont {Fujimoto}}]{Sato2009}%
  \BibitemOpen
  \bibfield  {author} {\bibinfo {author} {\bibfnamefont {M.}~\bibnamefont
  {Sato}}, \bibinfo {author} {\bibfnamefont {Y.}~\bibnamefont {Takahashi}}, \
  and\ \bibinfo {author} {\bibfnamefont {S.}~\bibnamefont {Fujimoto}},\ }\href
  {\doibase 10.1103/PhysRevLett.103.020401} {\bibfield  {journal} {\bibinfo
  {journal} {Phys. Rev. Lett.}\ }\textbf {\bibinfo {volume} {103}},\ \bibinfo
  {pages} {020401} (\bibinfo {year} {2009})}\BibitemShut {NoStop}%
\bibitem [{\citenamefont {Jiang}\ \emph {et~al.}(2011)\citenamefont {Jiang},
  \citenamefont {Kitagawa}, \citenamefont {Alicea}, \citenamefont {Akhmerov},
  \citenamefont {Pekker}, \citenamefont {Refael}, \citenamefont {Cirac},
  \citenamefont {Demler}, \citenamefont {Lukin},\ and\ \citenamefont
  {Zoller}}]{Jiang2011}%
  \BibitemOpen
  \bibfield  {author} {\bibinfo {author} {\bibfnamefont {L.}~\bibnamefont
  {Jiang}}, \bibinfo {author} {\bibfnamefont {T.}~\bibnamefont {Kitagawa}},
  \bibinfo {author} {\bibfnamefont {J.}~\bibnamefont {Alicea}}, \bibinfo
  {author} {\bibfnamefont {A.~R.}\ \bibnamefont {Akhmerov}}, \bibinfo {author}
  {\bibfnamefont {D.}~\bibnamefont {Pekker}}, \bibinfo {author} {\bibfnamefont
  {G.}~\bibnamefont {Refael}}, \bibinfo {author} {\bibfnamefont {J.~I.}\
  \bibnamefont {Cirac}}, \bibinfo {author} {\bibfnamefont {E.}~\bibnamefont
  {Demler}}, \bibinfo {author} {\bibfnamefont {M.~D.}\ \bibnamefont {Lukin}}, \
  and\ \bibinfo {author} {\bibfnamefont {P.}~\bibnamefont {Zoller}},\ }\href
  {\doibase 10.1103/PhysRevLett.106.220402} {\bibfield  {journal} {\bibinfo
  {journal} {Phys. Rev. Lett.}\ }\textbf {\bibinfo {volume} {106}},\ \bibinfo
  {pages} {220402} (\bibinfo {year} {2011})}\BibitemShut {NoStop}%
\bibitem [{\citenamefont {Bardyn}\ \emph {et~al.}(2012)\citenamefont {Bardyn},
  \citenamefont {Baranov}, \citenamefont {Rico}, \citenamefont {\ifmmode
  \dot{I}\else \.{I}\fi{}mamo\ifmmode~\breve{g}\else \u{g}\fi{}lu},
  \citenamefont {Zoller},\ and\ \citenamefont {Diehl}}]{Bardyn2012}%
  \BibitemOpen
  \bibfield  {author} {\bibinfo {author} {\bibfnamefont {C.-E.}\ \bibnamefont
  {Bardyn}}, \bibinfo {author} {\bibfnamefont {M.~A.}\ \bibnamefont {Baranov}},
  \bibinfo {author} {\bibfnamefont {E.}~\bibnamefont {Rico}}, \bibinfo {author}
  {\bibfnamefont {A.}~\bibnamefont {\ifmmode \dot{I}\else
  \.{I}\fi{}mamo\ifmmode~\breve{g}\else \u{g}\fi{}lu}}, \bibinfo {author}
  {\bibfnamefont {P.}~\bibnamefont {Zoller}}, \ and\ \bibinfo {author}
  {\bibfnamefont {S.}~\bibnamefont {Diehl}},\ }\href {\doibase
  10.1103/PhysRevLett.109.130402} {\bibfield  {journal} {\bibinfo  {journal}
  {Phys. Rev. Lett.}\ }\textbf {\bibinfo {volume} {109}},\ \bibinfo {pages}
  {130402} (\bibinfo {year} {2012})}\BibitemShut {NoStop}%
\bibitem [{\citenamefont {Diehl}\ \emph {et~al.}(2011)\citenamefont {Diehl},
  \citenamefont {Rico}, \citenamefont {Baranov},\ and\ \citenamefont
  {Zoller}}]{Diehl2011}%
  \BibitemOpen
  \bibfield  {author} {\bibinfo {author} {\bibfnamefont {S.}~\bibnamefont
  {Diehl}}, \bibinfo {author} {\bibfnamefont {E.}~\bibnamefont {Rico}},
  \bibinfo {author} {\bibfnamefont {M.~A.}\ \bibnamefont {Baranov}}, \ and\
  \bibinfo {author} {\bibfnamefont {P.}~\bibnamefont {Zoller}},\ }\href
  {http://dx.doi.org/10.1038/nphys2106} {\bibfield  {journal} {\bibinfo
  {journal} {Nat Phys}\ }\textbf {\bibinfo {volume} {7}},\ \bibinfo {pages}
  {971} (\bibinfo {year} {2011})}\BibitemShut {NoStop}%
\bibitem [{\citenamefont {Cooper}\ and\ \citenamefont
  {Shlyapnikov}(2009)}]{Cooper2009}%
  \BibitemOpen
  \bibfield  {author} {\bibinfo {author} {\bibfnamefont {N.~R.}\ \bibnamefont
  {Cooper}}\ and\ \bibinfo {author} {\bibfnamefont {G.~V.}\ \bibnamefont
  {Shlyapnikov}},\ }\href {\doibase 10.1103/PhysRevLett.103.155302} {\bibfield
  {journal} {\bibinfo  {journal} {Phys. Rev. Lett.}\ }\textbf {\bibinfo
  {volume} {103}},\ \bibinfo {pages} {155302} (\bibinfo {year}
  {2009})}\BibitemShut {NoStop}%
\bibitem [{\citenamefont {Nishida}(2009)}]{Nishida2009}%
  \BibitemOpen
  \bibfield  {author} {\bibinfo {author} {\bibfnamefont {Y.}~\bibnamefont
  {Nishida}},\ }\href {\doibase http://dx.doi.org/10.1016/j.aop.2008.10.011}
  {\bibfield  {journal} {\bibinfo  {journal} {Annals of Physics}\ }\textbf
  {\bibinfo {volume} {324}},\ \bibinfo {pages} {897 } (\bibinfo {year}
  {2009})}\BibitemShut {NoStop}%
\bibitem [{\citenamefont {Nishida}\ and\ \citenamefont
  {Tan}(2008)}]{Nishida2008}%
  \BibitemOpen
  \bibfield  {author} {\bibinfo {author} {\bibfnamefont {Y.}~\bibnamefont
  {Nishida}}\ and\ \bibinfo {author} {\bibfnamefont {S.}~\bibnamefont {Tan}},\
  }\href {\doibase 10.1103/PhysRevLett.101.170401} {\bibfield  {journal}
  {\bibinfo  {journal} {Phys. Rev. Lett.}\ }\textbf {\bibinfo {volume} {101}},\
  \bibinfo {pages} {170401} (\bibinfo {year} {2008})}\BibitemShut {NoStop}%
\bibitem [{\citenamefont {Christensen}\ \emph {et~al.}(2015)\citenamefont
  {Christensen}, \citenamefont {Levinsen},\ and\ \citenamefont
  {Bruun}}]{Christensen2015}%
  \BibitemOpen
  \bibfield  {author} {\bibinfo {author} {\bibfnamefont {R.~S.}\ \bibnamefont
  {Christensen}}, \bibinfo {author} {\bibfnamefont {J.}~\bibnamefont
  {Levinsen}}, \ and\ \bibinfo {author} {\bibfnamefont {G.~M.}\ \bibnamefont
  {Bruun}},\ }\href {\doibase 10.1103/PhysRevLett.115.160401} {\bibfield
  {journal} {\bibinfo  {journal} {Phys. Rev. Lett.}\ }\textbf {\bibinfo
  {volume} {115}},\ \bibinfo {pages} {160401} (\bibinfo {year}
  {2015})}\BibitemShut {NoStop}%
\bibitem [{\citenamefont {Heiselberg}\ \emph {et~al.}(2000)\citenamefont
  {Heiselberg}, \citenamefont {Pethick}, \citenamefont {Smith},\ and\
  \citenamefont {Viverit}}]{Heiselberg2000}%
  \BibitemOpen
  \bibfield  {author} {\bibinfo {author} {\bibfnamefont {H.}~\bibnamefont
  {Heiselberg}}, \bibinfo {author} {\bibfnamefont {C.~J.}\ \bibnamefont
  {Pethick}}, \bibinfo {author} {\bibfnamefont {H.}~\bibnamefont {Smith}}, \
  and\ \bibinfo {author} {\bibfnamefont {L.}~\bibnamefont {Viverit}},\ }\href
  {\doibase 10.1103/PhysRevLett.85.2418} {\bibfield  {journal} {\bibinfo
  {journal} {Phys. Rev. Lett.}\ }\textbf {\bibinfo {volume} {85}},\ \bibinfo
  {pages} {2418} (\bibinfo {year} {2000})}\BibitemShut {NoStop}%
\bibitem [{\citenamefont {Bijlsma}\ \emph {et~al.}(2000)\citenamefont
  {Bijlsma}, \citenamefont {Heringa},\ and\ \citenamefont
  {Stoof}}]{Bijlsma2000}%
  \BibitemOpen
  \bibfield  {author} {\bibinfo {author} {\bibfnamefont {M.~J.}\ \bibnamefont
  {Bijlsma}}, \bibinfo {author} {\bibfnamefont {B.~A.}\ \bibnamefont
  {Heringa}}, \ and\ \bibinfo {author} {\bibfnamefont {H.~T.~C.}\ \bibnamefont
  {Stoof}},\ }\href {\doibase 10.1103/PhysRevA.61.053601} {\bibfield  {journal}
  {\bibinfo  {journal} {Phys. Rev. A}\ }\textbf {\bibinfo {volume} {61}},\
  \bibinfo {pages} {053601} (\bibinfo {year} {2000})}\BibitemShut {NoStop}%
\bibitem [{\citenamefont {Abrikosov}\ \emph {et~al.}(1975)\citenamefont
  {Abrikosov}, \citenamefont {Gorkov},\ and\ \citenamefont
  {Dzyaloshinskii}}]{AGD}%
  \BibitemOpen
  \bibfield  {author} {\bibinfo {author} {\bibfnamefont {A.~A.}\ \bibnamefont
  {Abrikosov}}, \bibinfo {author} {\bibfnamefont {L.~P.}\ \bibnamefont
  {Gorkov}}, \ and\ \bibinfo {author} {\bibfnamefont {I.~E.}\ \bibnamefont
  {Dzyaloshinskii}},\ }\href@noop {} {\emph {\bibinfo {title} {Methods of
  Quantum Field Theory in Statistical Physics}}}\ (\bibinfo  {publisher} {Dover
  Publications},\ \bibinfo {year} {1975})\BibitemShut {NoStop}%
\bibitem [{\citenamefont {Anderson}\ and\ \citenamefont
  {Morel}(1961)}]{Anderson1961}%
  \BibitemOpen
  \bibfield  {author} {\bibinfo {author} {\bibfnamefont {P.~W.}\ \bibnamefont
  {Anderson}}\ and\ \bibinfo {author} {\bibfnamefont {P.}~\bibnamefont
  {Morel}},\ }\href {\doibase 10.1103/PhysRev.123.1911} {\bibfield  {journal}
  {\bibinfo  {journal} {Phys. Rev.}\ }\textbf {\bibinfo {volume} {123}},\
  \bibinfo {pages} {1911} (\bibinfo {year} {1961})}\BibitemShut {NoStop}%
\bibitem [{\citenamefont {Berezinskii}(1972)}]{Berezinskii1972}%
  \BibitemOpen
  \bibfield  {author} {\bibinfo {author} {\bibfnamefont {V.~L.}\ \bibnamefont
  {Berezinskii}},\ }\href@noop {} {\bibfield  {journal} {\bibinfo  {journal}
  {Soviet Physics JETP}\ }\textbf {\bibinfo {volume} {34}},\ \bibinfo {pages}
  {610} (\bibinfo {year} {1972})}\BibitemShut {NoStop}%
\bibitem [{\citenamefont {Kosterlitz}\ and\ \citenamefont
  {Thouless}(1973)}]{Kosterlitz1973}%
  \BibitemOpen
  \bibfield  {author} {\bibinfo {author} {\bibfnamefont {J.~M.}\ \bibnamefont
  {Kosterlitz}}\ and\ \bibinfo {author} {\bibfnamefont {D.~J.}\ \bibnamefont
  {Thouless}},\ }\href {http://stacks.iop.org/0022-3719/6/i=7/a=010} {\bibfield
   {journal} {\bibinfo  {journal} {Journal of Physics C: Solid State Physics}\
  }\textbf {\bibinfo {volume} {6}},\ \bibinfo {pages} {1181} (\bibinfo {year}
  {1973})}\BibitemShut {NoStop}%
\bibitem [{\citenamefont {Kosterlitz}(1974)}]{Kosterlitz1974}%
  \BibitemOpen
  \bibfield  {author} {\bibinfo {author} {\bibfnamefont {J.~M.}\ \bibnamefont
  {Kosterlitz}},\ }\href {http://stacks.iop.org/0022-3719/7/i=6/a=005}
  {\bibfield  {journal} {\bibinfo  {journal} {Journal of Physics C: Solid State
  Physics}\ }\textbf {\bibinfo {volume} {7}},\ \bibinfo {pages} {1046}
  (\bibinfo {year} {1974})}\BibitemShut {NoStop}%
\bibitem [{\citenamefont {Jos\'e}\ \emph {et~al.}(1977)\citenamefont {Jos\'e},
  \citenamefont {Kadanoff}, \citenamefont {Kirkpatrick},\ and\ \citenamefont
  {Nelson}}]{Jose1977}%
  \BibitemOpen
  \bibfield  {author} {\bibinfo {author} {\bibfnamefont {J.~V.}\ \bibnamefont
  {Jos\'e}}, \bibinfo {author} {\bibfnamefont {L.~P.}\ \bibnamefont
  {Kadanoff}}, \bibinfo {author} {\bibfnamefont {S.}~\bibnamefont
  {Kirkpatrick}}, \ and\ \bibinfo {author} {\bibfnamefont {D.~R.}\ \bibnamefont
  {Nelson}},\ }\href {\doibase 10.1103/PhysRevB.16.1217} {\bibfield  {journal}
  {\bibinfo  {journal} {Phys. Rev. B}\ }\textbf {\bibinfo {volume} {16}},\
  \bibinfo {pages} {1217} (\bibinfo {year} {1977})}\BibitemShut {NoStop}%
\bibitem [{\citenamefont {Taylor}\ \emph {et~al.}(2006)\citenamefont {Taylor},
  \citenamefont {Griffin}, \citenamefont {Fukushima},\ and\ \citenamefont
  {Ohashi}}]{Taylor2006}%
  \BibitemOpen
  \bibfield  {author} {\bibinfo {author} {\bibfnamefont {E.}~\bibnamefont
  {Taylor}}, \bibinfo {author} {\bibfnamefont {A.}~\bibnamefont {Griffin}},
  \bibinfo {author} {\bibfnamefont {N.}~\bibnamefont {Fukushima}}, \ and\
  \bibinfo {author} {\bibfnamefont {Y.}~\bibnamefont {Ohashi}},\ }\href
  {\doibase 10.1103/PhysRevA.74.063626} {\bibfield  {journal} {\bibinfo
  {journal} {Phys. Rev. A}\ }\textbf {\bibinfo {volume} {74}},\ \bibinfo
  {pages} {063626} (\bibinfo {year} {2006})}\BibitemShut {NoStop}%
\bibitem [{\citenamefont {Fisher}\ \emph {et~al.}(1973)\citenamefont {Fisher},
  \citenamefont {Barber},\ and\ \citenamefont {Jasnow}}]{Fisher1973}%
  \BibitemOpen
  \bibfield  {author} {\bibinfo {author} {\bibfnamefont {M.~E.}\ \bibnamefont
  {Fisher}}, \bibinfo {author} {\bibfnamefont {M.~N.}\ \bibnamefont {Barber}},
  \ and\ \bibinfo {author} {\bibfnamefont {D.}~\bibnamefont {Jasnow}},\ }\href
  {\doibase 10.1103/PhysRevA.8.1111} {\bibfield  {journal} {\bibinfo  {journal}
  {Phys. Rev. A}\ }\textbf {\bibinfo {volume} {8}},\ \bibinfo {pages} {1111}
  (\bibinfo {year} {1973})}\BibitemShut {NoStop}%
\bibitem [{\citenamefont {Lieb}\ \emph {et~al.}(2002)\citenamefont {Lieb},
  \citenamefont {Seiringer},\ and\ \citenamefont {Yngvason}}]{Lieb2002}%
  \BibitemOpen
  \bibfield  {author} {\bibinfo {author} {\bibfnamefont {E.~H.}\ \bibnamefont
  {Lieb}}, \bibinfo {author} {\bibfnamefont {R.}~\bibnamefont {Seiringer}}, \
  and\ \bibinfo {author} {\bibfnamefont {J.}~\bibnamefont {Yngvason}},\ }\href
  {\doibase 10.1103/PhysRevB.66.134529} {\bibfield  {journal} {\bibinfo
  {journal} {Phys. Rev. B}\ }\textbf {\bibinfo {volume} {66}},\ \bibinfo
  {pages} {134529} (\bibinfo {year} {2002})}\BibitemShut {NoStop}%
\bibitem [{foo({\natexlab{a}})}]{footnote2}%
  \BibitemOpen
  \href@noop {} {\bibfield  {journal} {\bibinfo  {journal} {In principle the
  gap equation should be solved together with the particle number equation.
  However, the change of chemical potential due to paring is negelible for our
  calculations}\ }}\BibitemShut {NoStop}%
\bibitem [{foo({\natexlab{b}})}]{footnote}%
  \BibitemOpen
  \href@noop {} {\bibfield  {journal} {\bibinfo  {journal} {For example, the
  lowest temperature that can be achieved for a 3D Fermi gas at the ENS lab is at
  least $\ve_F/20$ (private communiation with F. Chevy)}\ }}\BibitemShut {NoStop}%
\end{thebibliography}

%

\widetext
\clearpage
\begin{center}
\textbf{\large Supplemental Material }\\
\vspace{4mm}
{Zhigang Wu and G. M. Bruun}\\
\vspace{2mm}
{\em \small
Department of Physics and Astronomy, Aarhus University, DK-8000 Aarhus C, Denmark
}\end{center}

\section{The $p$-wave component of the induced interaction}
To calculate $V_1(p,q;i \omega_\nu)$ defined in Eq.~(\ref{V1}) we first introduce the dimensionless quantities $ \bar q = \bar q/k_{F}$ and $\bar \omega_\nu = \omega_\nu/ \ve_{F}$, where $\ve_F = k_F^2/2m_F$ is the Fermi energy of the 2D gas.   In terms of these quantities, $\calN_FV_{\rm ind}(\bar\bp - \bar\bq,i\bar \omega_\nu)$, where $\calN_F = m_F/2\pi$ is the density of states of the Fermi gas, can be written as
 \begin{align}
 \calN_F V_{\rm ind}(\bar\bp - \bar\bq,i\bar\omega_\nu)& = -\sqrt{\pi}\left(1+\frac{m_F}{m_B}\right ){\beta_{FB}^2}\left \{\frac{1-{1}/{\sqrt{1-r_e^2\bar \omega_\nu^2}}}{\sqrt{2\beta_{BB}\left (1- \sqrt{1-r_e^2\bar \omega_\nu^2}\right )+r_{FB}^2\left [\bar p^2+\bar q^2-2\bar p\bar q\cos(\phi_{\bar \bp}-\phi_{\bar \bq})\right]}} \right. \nn \\
 &\left. + \frac{1+{1}/{\sqrt{1-r_e^2\bar \omega_\nu^2}}}{\sqrt{2\beta_{BB}\left (1+\sqrt{1-r_e^2\bar \omega_\nu^2}\right )+r_{FB}^2\left [\bar p^2+\bar q^2-2\bar p\bar q\cos(\phi_{\bar \bp}-\phi_{\bar \bq})\right]}} \right \}
  \label{Vind}
 \end{align}
 where $\beta_{FB}\equiv  (n_Ba_{\rm eff}^3)^{1/3}$, $\beta_{BB}\equiv  (n_Ba_B^3)^{1/3}$,  $r_{FB}\equiv n_F^{1/2}/n_B^{1/3}$ and $r_e \equiv \e_{F}/g_Bn_B = ({m_B}/{m_F}){r_{FB}}/{\beta_{BB}}$.  We see from the above expression that aside from the simple dependence on $\beta_{FB}$, the overall strength of the induced interaction is determined by three additional parameters, $r_{FB}$, $\beta_{BB}$ and ${m_F}/{m_B}$. 
 
 Substituting Eq.~(\ref{Vind}) into Eq.~(\ref{V1}) we find
 \begin{align}
 \calN_F V_1(\bar p,\bar q;i\bar \omega_\nu) & = -\sqrt{\frac{2}{\pi}}\left(1+\frac{m_F}{m_B}\right )\frac{\beta_{FB}^2}{r_{FB}\sqrt{\bar p\bar q}}\left [\left( 1-\frac{1}{\sqrt{1-r_e^2\bar \omega_\nu^2}}\right )\frac{f_- K\left(\frac{2}{1+f_-}\right)-(1+f_-)E\left(\frac{2}{1+f_-}\right)}{\sqrt{1+f_-}} \right. \nn \\
 &\left. + \left( 1+\frac{1}{\sqrt{1-r_e^2\bar \omega_\nu^2}}\right ) \frac{f_+K\left(\frac{2}{1+f_+}\right)-(1+f_+)E\left(\frac{2}{1+f_+}\right)}{\sqrt{1+f_+}}\right],
 \label{NV1f}
 \end{align}
where
 \beq
 f_\pm  = \frac{\beta_{BB}}{r^2_{BF}\bar p \bar q}\left (1\pm\sqrt{1-r_e^2\bar \omega_\nu^2}\right )+\frac{\bar p^2+\bar q^2}{2\bar p\bar q},
 \eeq
 and $K(x)$ and $E(x)$ are the complete elliptic integral of the first and second kind. We note that for $|r_e\bar \omega_\nu| > 1$ the two terms inside the brackets of Eq.~(\ref{NV1f}) are complex conjugates of each other so that $\calN_F V_1(\bar p,\bar q;i\bar\omega_\nu) $ is always real. The zero frequency component of the expression in Eq.~(\ref{NV1f}) at the Fermi surface is plotted in Fig.~\ref{NV}.

 \section{Frequency dependent gap equation and superfluid density}
With the $p$-wave form of the gap parameter $ \Delta(\bp,i\omega_n) = \Delta_1(p,i\omega_n)e^{i\phi_\bp}$, the gap equation (\ref{gapeq_fre}) reduces to 
 \begin{align}
 \Delta_1(p,i\omega_n) &= -\frac{1}{2\pi}T\sum_m\int{dq q}V_1(p,q;i\omega_n-i\omega_m) \frac{\Delta_1(q,i\omega_m)}{\omega_m^2+\calE(\bq,i\omega_m)^2} 
  \label{gapeq1}
 \end{align}
 
The mean-field critical temperature is determined by the linearised gap equation 
 \begin{align}
 \Delta_1(p,i\omega_n) &= -\frac{1}{2\pi}T\sum_m\int{dq q}V_1(p,q;i\omega_n-i\omega_m) \frac{\Delta_1(q,i\omega_m)}{\omega_m^2+\xi_\bq^2} \nn \\
  & \equiv \hat L\left ( \Delta_1\right ),
  \label{gapeq2}
 \end{align}
 where $\hat L$ denotes the integral operator in the above equation. The mean-field critical temperature is obtained from the condition that the largest eigenvalue of $\hat L$ becomes unity.
 
To determine the superfluid density we consider the grand potential of a flowing superfluid 
\beq
\Omega_{\rm MF} (\bv,T,\mu) = \frac{T}{\calA} S_{\rm eff}(\bv) = - \frac{T}{\calA} \ln \calZ_{\rm MF} (\bv) ,
\eeq
where
\begin{align}
 \calZ_{\rm MF} (\bv) 
 &= {\rm{det} \left (-{2\beta \calA}V_{\rm ind}^{-1} \right ) }  \exp \left \{ \frac{\beta \calA}{2}\sum_{p,p'}\Delta^*(p)V_{\rm ind}^{-1}(p-p')\Delta(p)\right \}  \int \calD (\bar a, a )\exp \left \{ -\sum_{\omega_n >0,\bp} \bar \Lambda(p) G^{-1}(p,\bq) \Lambda(p) \right \},
 \end{align}
 where $\bq = m_F\bv$, $\bar \Lambda (p) = [\bar a(p)\,\, a(-p)]$ and $\Lambda (p) = [a(p)\,\, \bar a(-p)]^{\rm T}$. Here $G^{-1}(p,\bq)$ is given by
\begin{align}
G^{-1}(p,\bq) = \begin{bmatrix}
    -i\omega_n +\xi_{\bp+\bq} e^{i\omega_n0^+}   &\Delta(p) \\
     \Delta^*(p)    & -i\omega_n -\xi_{-\bp+\bq} e^{-i\omega_n0^+} \\
\end{bmatrix}. 
\end{align}
One finds
 \begin{align}
  \Omega_{\rm MF}(\bv,T,\mu) & = const. -\frac{1}{2}\sum_{p,p'}\Delta^*(p)V_{\rm ind}^{-1}(p-p')\Delta(p) \nn \\
  &-\frac{T}{2\calA}\sum_{p} \ln \left [  -\left(-i\omega_n+ \xi_{\bp}e^{i\omega_n0^+}\right )\left(-i\omega_n-\xi_{\bp}e^{-i\omega_n 0^+}\right )  + |\Delta(p)|^2  - R(v) \right ]  \label{dOmega2}
 \end{align}
 where
  \beq 
 R(v)  = -\frac{1}{4}m_F^2v^4+(\bp\cdot\bv)^2 -m_Fv^2\xi_\bp -\frac{1}{2}i\omega_n m_F v^2 \left(e^{i\omega_n 0^+}-e^{-i\omega_n 0^+}\right ) -i\omega_n \bp\cdot\bv  \left(e^{i\omega_n 0^+} + e^{-i\omega_n 0^+}\right ) 
 \eeq
Using $\rho_s = \partial_v^2\Omega_{\rm MF}(\bv)|_{\bv=0}$ together with Eq.~(\ref{dOmega2}) we arrive at the expression in Eq.~(\ref{ns}).
\end{document}